# Why is Open Access Development so Successful? Stigmergic organization and the economics of information

Francis Heylighen

ECCO, Vrije Universiteit Brussel
http://pcp.vub.ac.be/HEYL.html

**Abstract**: The explosive development of "free" or "open source" information goods contravenes the conventional wisdom that markets and commercial organizations are necessary to efficiently supply products. This paper proposes a theoretical explanation for this phenomenon, using concepts from economics and theories of self-organization. Once available on the Internet, information is intrinsically not a scarce good, as it can be replicated virtually without cost. Moreover, freely distributing information is profitable to its creator, since it improves the quality of the information, and enhances the creator's reputation. This provides a sufficient incentive for people to contribute to open access projects. Unlike traditional organizations, open access communities are open, distributed and self-organizing. Coordination is achieved through stigmergy: listings of "work-in-progress" direct potential contributors to the tasks where their contribution is most likely to be fruitful. This obviates the need both for centralized planning and for the "invisible hand" of the market.

## 1. Introduction

The last few years have witnessed a surprisingly quick and successful spread of "free", "libre", "open access", "open content" or "open source" information products. In this paper, I will skip the interesting but sometimes complicated differences between the precise forms of distribution that are associated with these terms, and focus on their common characteristic, which is that these products are not *proprietary*: they are not "owned" by a particular individual or organization who has the sole rights to distribute them. They are a part of the "creative commons" that everyone can freely access, use, and—in many cases—modify. I will also ignore the differences between different media or intended applications (such as text, images, music, or software), but focus on their common characteristic, which is that they consist purely of information that can be duplicated without limit. As shorthand, I will designate them all as "open access".

Complex software applications, websites, journals and magazines, books, pictures, "podcasts", video recordings and even whole encyclopedias have been made accessible by their authors to everyone, for consultation, use and even modification, and this without cost or restriction. Perhaps the best known examples are the Linux[1] operating system (Moody, 2002), which is starting to compete with Microsoft Windows to become the standard on which computers run, Wikipedia[2], an international web encyclopedia, that is already the largest one in existence (Lih, 2004), ArXiv[3] and other Internet paper archives where thousands of scientists make their results freely available before they are published in proprietary journals (Hajjem et al. 2005), and finally the World-Wide Web itself, a collection of communication protocols and software applications for the transparent distribution of hypermedia documents across the Internet (Berners-Lee, 1999).

These developments are revolutionizing our society. On the one hand, they put into question one of the foundations of the present-day market economy, the idea that intellectual property is necessary to stimulate innovation. On the other hand, they open up huge opportunities, which include: freely providing software, technical know-how, scientific knowledge and general education to the countries and people that need it most, but can least afford to pay for it; empowering and stimulating ordinary people to be intellectually creative and thus help others; reducing the danger of commercial monopolies that control software standards or news distribution; and creating and distributing information much more quickly and widely than before, when it is needed and where it is needed.

While this development clearly rests on the Internet, it was largely overlooked in the recent past of the 'dot-com bubble', when pundits were focusing on the great opportunities for commercial exploitation of the net (Howcroft, 2001). With the advent of the web and its multimedia capabilities for information delivery, most experts were wondering how the big corporations such as ABC or Time-Warner would be able to produce enough "content" to satisfy the huge surge in demand for information that the web would bring along. Presently, the web users themselves are producing the "content", in millions of blogs and wikis, providing news, opinion, entertainment and information, at a fraction of the cost that the corporations were planning to invest.

This development was less surprising to the pioneers of the World-Wide Web (among which I count myself, having developed the complex Principia Cybernetica website[4] already in 1993, cf. Heylighen, 1994). Before the advent of commercial interests on the Internet, the reigning culture among its users was one of freedom, cooperation and sharing, not of competition and exclusion. The early Internet users were mostly researchers, who found it obvious that they would make the results of their efforts publicly available, without demanding any money. In that, they merely followed the basic philosophy that has made science the most important driver of social, technological and economical progress: publish your data and ideas as widely as possible, so that others can use them, criticize them, and improve on them. Before 1996, basically all information and software on the web was free, but that did not seem to deter its producers from being creative, or from releasing additional material and improved versions at a breakneck pace.

---

[1] http://www.linux.org
[2] http://www.wikipedia.org
[3] http://www.arxiv.org
[4] http://pespmc1.vub.ac.be

This free spirit was eclipsed by the dot-com boom ("com" standing for "commercial"), where businesses turned en masse to the Internet in the hope of making money via advertising, sales or intellectual property rights (Howcroft, 2001). The burst of the bubble in 2001 made it clear that moneymaking on the Internet was not as easy as people expected. Part of the reason was simply that there was already so much available on the web for free: why would you take a complex and expensive subscription to the Encyclopaedia Britannica website if you could freely access most of that information elsewhere? Of course, that doesn't mean that the Internet cannot be used for commercial transactions, as exemplified by highly successful businesses such as Amazon.com and eBay. But these function basically as intermediaries for the sales of traditional material goods, rather than by selling pure information.

This evolution suggests that there is something special about the distribution of information via the Internet, which contravenes conventional wisdom about economics. This paper intends to explore the deeper mechanisms underlying the success of open access development and distribution, from the perspective of complex, evolving systems. But first we need to analyze in more detail why this phenomenon was so surprising to most.

## 2. Economic theory and open access development

The classical economics model holds that people are intrinsically selfish, and will not do anything to help others—such as providing information products—without remuneration. Traditional economics is based on the assumption that private property rights are needed as an incentive for production. Only when you have full control over your production can you ask remuneration for it to the people who would like to use it. Moreover, the free market model assumes that competition is needed to optimize production: if people do not buy your products because they prefer the one of your competitors, you will be forced to improve your products or lower their price. When the major producers all cooperate, as in a cartel, competition is eroded and prices can increase freely without corresponding increase in quality.

At first sight, all these economic principles are contravened by the open access community (cf. Lerner & Tirole, 2002): people produce information or software for free, allow others to use it as they please, and all work together on major enterprises such as Linux or Wikipedia. Yet, this community has produced better products, in a shorter time span, and at a lower cost, than specialized private enterprises that have been in the business for decades.

The paradox becomes even greater when we consider the aspects of organization and control. According to theory, the major advantage of a free market economy over a plan-based economy is coordination: a centralized planning institute can never collect and process all the information needed to decide what to produce when and where; a free market, on the other hand, functions like an "invisible hand", that automatically allocates the right amount of resources to the production of each commodity that is needed (Hayek, 1945). This happens through the *law of supply and demand* and the *price mechanism*: whenever demand for a good exceeds supply, its price will increase, thus enticing suppliers to produce more of that good. In that way, demand and supply are automatically balanced by a negative feedback control mechanism, without need for complicated planning.

Institutional economics adds a qualification to the power of market mechanisms, though. Its main assumption is that individuals competing in a market will start to formally cooperate, thus forming an organization or firm, in order to reduce *transaction costs* (Williamson & Maste, 1995). Transaction costs arise because of the need to negotiate and arrange any exchange of goods, services or money. These negotiations cost a lot of effort and time, while failing to fully eliminate basic uncertainties (is this product or service reliable? are there loopholes in the contract?). A firm is based on a set of agreed-upon rules governing the interactions between its employees, so as to minimize negotiations and uncertainties. This requires drawing a clear boundary between those who belong to the organization (and therefore are expected to obey the rules) and those who do not. That also helps to ensure that no private information is leaked out of the group, where it could be exploited by competitors.

Whereas regulations may be more reliable than a free market, they are less flexible: they cannot adapt automatically to new circumstances. Therefore, the organization requires a form of intelligent management to coordinate the activities of the employees, and direct them to the most important task at hand. This is normally achieved by means of a hierarchical structure, with a CEO or board of directors at the top, to plan and control the activity and issue commands down the line to the lower levels. This control system moreover must ensure that the employees play according to the rules, and do not selfishly exploit the opportunities without giving their due effort in return. In other words, the management must thwart the ever-present danger of "free riders" by implementing effective punishment when necessary.

The paradox of the open access community is that it seems to ignore most of these organizational principles as well. In general, anyone can join or leave a given community at any moment, and there are no formal members or employees, as different people tend to be involved to different degrees. Moreover, the community is typically decentralized, without formal, hierarchical structure or punishments for free riders. Raymond (1999) has called this loose, self-organizing cooperation model a "bazaar", as contrasted with the "cathedral" model that exemplifies closed, centralized, hierarchical organizations. Yet, this distributed coordination cannot rely on the market mechanism either, since there are no prices for products to signal where demand is greatest.

In conclusion, open access development not only contravenes common business wisdom, but some of the most fundamental assumptions of economic theory. This means that we need to develop an alternative theory to explain how open access can function.

## 3. Incentives for information sharing

A first essential property to note is the non-material nature of information. Information is not subjected to the physical constraint of the conservation of matter and energy, and therefore to the economic constraint of *scarcity*: once you have gotten a piece of information, such as a computer program, you can multiply and distribute it without limit, at virtually no cost. Giving away these copies to others does not deprive the original owner. The fact that you use a particular piece of information does not in any way preclude somebody else from using the same information at the same time. This property is called *non-rivalry* in economics (e.g. Martens, 2004). While it puts the fundamental economic assumption of scarcity on its head, economic theory has as yet little to say about how to deal with non-rival goods (De Long & Froomkin, 1998).

Another important property of information is its *partial excludability*: while you could in principle exclude somebody else from using the same information, e.g. via copyright law, patenting or copy protection, the easy replication of information makes this prohibition increasingly difficult to implement. This is a problem for traditional economics, which assumes that the producer of a good, whether rival or not, needs an incentive to continue producing it, and for that must have some means to enforce consumers of that good to pay for it (Martens, 2004; De Long & Froomkin, 1998).

Yet, in open access communities we see a very different structure of incentives. First, the non-rivalry of information explains why open access development can function with relatively few incentives: assuming that you have the needed resources (hardware, software, expertise, time...) to produce a piece of information that you want for yourself (e.g. a program, a bibliography, a poem, a photo of your dog…), then it hardly costs you more to also make it available to others. Thus, a minor investment of your own time such as a hobby, an accidental discovery or a quick hack, can nevertheless produce an information good that benefits thousands (cf. Ghosh, 1998).

Although the direct effect is likely to be tiny, this benefit to others may indirectly benefit you. Indeed, if the people in your community become a little more efficient, productive, or simply happy, because of something you contributed then your own life in the community will become better, even if nobody knows that it was you who contributed. Thus, the idea of contributing to the community will appeal to the same instinct of altruism or "goodness" that makes people give money to charity or do volunteer work. This is a first incentive for sharing your information products (Weber, 2004). However, while sociobiology and evolutionary psychology (Buss, 1995) have shown altruism to be a true, genetically based motive, on its own it does not seem to be strong enough to support a complex economy, as shown by the *free rider problem* (cf. Heylighen, 2007) and the failure of communism.

First, we must note that the classical free rider problem does not exist for non-rival resources. A *free rider* is defined as an individual who profits from investments made by others but without doing an equivalent effort in return. Most users of open access information fall under that category: they utilize products made by others, but contribute little or nothing themselves. This apparent "parasitism" or non-reciprocity in this case is not a problem, though: the producers do not benefit less from the information they produced because others access it too. This already explains why open access communities lack the typical controls and penalties that social groups throughout history have evolved to discourage profiteering. Moreover, as we are about to show, the producers receive benefits that the mere "consumers" fail to get, and as such maintain a competitive advantage over free riders.

So, let us proceed to more selfish motivations for making your work open access. The minimal effort of sharing your results with others will be more than compensated for by the fact that these others may in turn produce or suggest improvements that benefit the creator. For example, your interest in a particular historical figure or geographic location may lead you to write down what you know about the subject in a Wikipedia article. This article is likely to incite others to add details that you weren't aware of, thus in turn helping you advance. Moreover, expressing it in an explicit, public form is likely to provide you with *feedback* (which can be as little as someone saying "nice work", or you noticing that the program isn't used quite as expected). Feedback ("reinforcement") is the basic driver of learning. Thus, even if your actual product is not improved by publicizing it, your expertise in producing it may well be.

If the opportunity of getting a better product and becoming better yourself is not enough of an incentive, there is an even stronger motivation for contributing to the collective. Indeed, doing so makes you visible within the community, earning you *recognition* for your expertise, activity, and altruism. Psychologists have proposed that earning esteem*,* status, or a good reputation within your community is a fundamental human drive (Maslow, 1970). From the point of view of evolutionary psychology (Buss, 1995), it is probably even more basic than gaining wealth, since during most of (pre)history the person with the highest status in the group would anyway get the best access to material resources—in addition to other resources, such as mates, friends, help, or information. Surveys have shown that the development of a good reputation in the field is indeed a concrete incentive for many open source developers (Lerner & Tirole, 2002).

One of the ways in which these developers can still earn a conventional income is by giving away their information products (e.g. software or blogs) for free, but demanding a fee for consultancy. Indeed, once complex software is adopted by many users, some of these may be willing to pay for help with specific problems by consulting the only true experts, namely the ones who wrote the software. Creators of literary or artistic content may adopt similar methods. For example, a rock group may give away their recordings for free, but once they have established sufficient popularity in that way, they can charge people for attending their concerts. Similarly, authors or journalists may provide free access to their texts (e.g. blogs), but make money by charging for interviews or lectures given to a restricted audience.

This way of earning money makes more sense than selling access to proprietary information, since it focuses on the one thing that effectively becomes more scarce in an information society: personal *attention* (Simon, 1971). Whatever amount of information a person makes freely accessible, an important part of his or her expertise still remains tacit (Reber, 1993), and therefore not available for easy replication. Such intuitive knowledge will only come to the surface when it is used to tackle a particular problem or question. Since a person has only a limited time, energy or attention to spend on any particular issue, this implicit knowledge does *not* obey the property of non-rivalry or unlimited replicability that characterizes explicit information. Therefore, the laws of economics dictate that people will be willing to pay in order to get access to the personal attention of an expert.

However, if you want to sell your advice to others for good money, you will have to convince them of your authority in the matter. Here reputation becomes a major factor. An even more direct way to get recognition for your expertise is by making its products publicly (and preferably freely) available, so that people can judge for themselves how good you are. This is much more difficult if you work for a firm, since outsiders can only get a dim view of who contributed what within a closed organization (cf. Lerner & Tirole, 2002). A similar mechanism is already used in the academic world: scientists' level of expertise is judged primarily by the quantity and quality of their *publications*, while their eventual income is largely dependent on that perceived level of expertise. Research has shown that making a publication open access increases the number of citations it gets, which is the most direct measure of the visibility and reputation of its author (Hajjem, Harnad & Gingras, 2005).

What this system of remuneration still lacks is a way to pay for the initial investments that have to be made before one can start to produce useful information. This includes hardware, software, Internet access, and education. Happily, all these factors are quickly declining in price thanks to on-going progress in information technology. The most expensive part, basic education, is provided (nearly) for free in most countries, as the burden is being shouldered by society as a whole—which has long ago understood that a well-educated workforce is to everybody's benefit. More advanced education typically remains expensive, although scholarships tend to be available for those who show most promise (e.g. in the work they have already published). But here again, advances in information production, distribution and the open access philosophy that tends to go with it, hold the promise that soon educational material for any domain will be available for free on the Internet[5].

## 4. Self-organization through stigmergy

To understand the distributed organization that characterizes open access development, we can draw inspiration from recent theories of self-organization (e.g. Heylighen & Gershenson, 2003) and complex adaptive systems (e.g. Muffatto & Faldani, 2003). A particularly relevant idea, used in the modelling of collective intelligence (Heylighen, 1999) and the simulation of swarming behavior (Bonabeau, Dorigo & Theraulaz, 1999), is the concept of *stigmergy* (Susi & Ziemke, 2001). A process is stigmergic if the work ("*ergon*" in Greek) done by one agent provides a stimulus ("*stigma*") that entices other agents to continue the job.

This concept was initially proposed to explain how a "bazaar" of dumb, uncoordinated termites manage to build their complex, "cathedral-like" termite hills (Grassé, 1959). The basic idea is that a termite initially drops a little bit of mud in a random place, but that the heaps that are formed in this way stimulate other termites to add to them (rather than start a heap of their own), thus making them grow higher until they touch other similarly constructed columns. The termites do not communicate about who is to do what how or when. Their only communication is indirect: the partially executed work of the ones provides information to the others about where to make their own contribution. In this way, there is no need for a centrally controlled plan, workflow, or division of labor.

While people are of course much more intelligent than social insects and do communicate, open access development uses essentially the same stigmergic mechanism (cf. Elliott, 2006, and the simulation by Robles et al., 2005): any new or revised document or software component uploaded to the site of a community is immediately scrutinized by the members of the community that are interested to use it. When one of them discovers a shortcoming, such as a bug, error or lacking functionality, that member will be inclined to either solve the problem him/herself, or at least point it out to the rest of the community, where it may again entice someone else to take up the problem.

---

[5] see e.g. http://www.wikibooks.org and http://www.globaltext.org

Like stigmergic organization in insects (Bonabeau et al., 1999), the process is self-reinforcing or autocatalytic (Heylighen, 1999; Heylighen & Gershenson, 2003): the more high quality material is already available on the community site, the more people will be drawn to check it out, and thus the more people are available to improve it further. Thus, open access can profit from a *positive feedback cycle* that boosts successful projects. This explains the explosive growth of systems such as Wikipedia or Linux. (A possible disadvantage of such "rich get richer" dynamics is that equally valuable, competing projects, because of random fluctuations or sequence effects, may fail to get the critical mass necessary to "take off".)

While most large-scale open access projects (such as Linux) have one or a few central figures (such as Linus Torvalds) that determine the general direction in which the project is headed, this control is much less strict than in traditional hierarchical organizations. Most of the work is typically performed in a distributed, self-organizing way. The lack of precise planning is more than compensated for by the fact that information about the present state of the project is completely and freely available, allowing anyone to contribute to anything at any moment. This provides for a much larger diversity of perspectives and experiences that are applied to finding and tackling problems, resulting in what Raymond (1999) calls *Linus' Law*: "given enough eyeballs, all bugs are shallow". Moreover, since contributors select the tasks they work on themselves, they tend to be more interested, motivated and knowledgeable about these tasks.

In this way, open access development fully profits from the evolutionary dynamic of *variation*, *recombination* and *selection* (van Wendel de Joode, 2004; Muffatto & Faldani, 2003). Openness attracts a greater number and diversity of participants, increasing the likeliness of cross-fertilization of their ideas into new combinations. This strongly accelerates the variation that is necessary to produce evolutionary novelty. This large and diverse community moreover enhances selection, since the new ideas will be tested in many more different circumstances, thus systematically eliminating the errors and weaknesses that might not have shown in a more homogeneous environment. All in all, this leads to greater flexibility, innovation, and reliability.

Stigmergy is more than blind variation and natural selection, though: the visible traces of the work performed previously function as a *mediator* system (Heylighen, 2007), storing and (indirectly) communicating information for the community. In that way, the mediator coordinates further activity, directing it towards the tasks where it is most likely to be fruitful. This requires a *shared workspace* accessible to all contributors (similar to what in AI is called a "blackboard system"). This external memory registers which tasks have already been performed and what problems still need to be tackled. The Web has provided a very powerful such workspace, since it enables the storage and public sharing of any "work-in-progress" information product.

To better understand the methods used by open access communities, we need to further distinguish direct from indirect stigmergy. In direct stigmergy, as exemplified by the termite-hill building, it is the "work-in-progress" itself that directs subsequent contributions. Indirect stigmergy may be exemplified by the way ants create trails of pheromones that direct other ants to food sources. The trails are left as "side-effects" of the actual work being performed: finding and bringing food to the nest. Such specially created traces may be needed because the task—finding the proverbial "needle (food) in the haystack (surroundings)"—is too complex to be performed without detailed clues. Thus, "indirect stigmergy" uses an additional medium for information storage. Yet, the coordination achieved in this way still keeps the hallmarks of distributed self-organization: the information is addressed to no one in particular, and may or may not be picked up by a particular individual at a particular moment.

In open access development, indirect stigmergy can be recognized in forums where bugs or feature requests are posted. These forums are themselves not part of the information product being developed, but they are regularly consulted by the developers, thus attracting their attention to tasks that seem worth performing. The problem with such an additional medium is that it adds to the complexity of the (self-)organization, especially if there is a lot of potentially relevant information posted there so that it becomes difficult to establish priorities. Here again we can learn a lesson from social insects. The pheromone trails left by ants undergo an efficient form of reinforcement learning (Heylighen, 1999): trails that lead to rich food sources will be used by many ants and thus amplified, trails that lead to poor or empty sources will weaken and eventually disappear. Since ants preferentially follow strong trails, this mechanism ensures that the most important tasks or opportunities are tackled first.

Applied to open access development, this means that we need adaptive mechanisms to make the most important requests stand out. An example of such mechanism can be found in Wikipedia. When a contributor marks a word as a hyperlink, but there is no article discussing this concept yet, an empty page is created inviting other contributors to fill in its content. This is direct stigmergy: whenever people look for that concept, they are directed immediately to the work that still needs to be done. But when there are many thousands of as yet incomplete entries, priorities must be established. Rather than having a central committee decide which entries are most important, Wikipedia implements a simple form of collective decision-making (cf. Heylighen, 1999): the entries that have most hyperlinks pointing to them are listed first in an automatically generated list of "most wanted articles".

Such a mechanism to display collective demand can be seen as the non-proprietary analog of the market. The price mechanism efficiently allocates resources to the production of those goods for which demand is highest, by offering the highest monetary rewards for them. Similarly, a "most wanted" ordering of requests offers the highest probability of recognition for the work performed, or of "good feelings" engendered by an altruistic deed. Such stigmergic prioritization is arguably even more efficient than a market, since there is no need for the complex and often irrational processes of buying, selling, bargaining and speculation that determine the eventual price of a commodity—leading to the typically chaotic movements of commodity prices on the stock exchange. Moreover, although price could be interpreted as a—very abstract—stigmergic signal, this variable is merely one-dimensional. On the other hand, open access tasks could be ranked on a website according to independent stigmergic criteria, such as urgency, difficulty, expected utility, required expertise, etc. In that way, potential contributors would be helped in finding the task that suits them best.

# 5. Conclusion

Since the fall of the Soviet Union, the common assumption has been that markets, private property rights, and commercial organizations are necessary to efficiently produce and distribute products. In addition to the apparent failure of communism, this view has been supported by two centuries of economic thought developing sophisticated models that purport to show that the market is the optimal way to allocate resources. In the last few years, however, the collective development of "open access" information products on the Web has emerged like a salient exception to this conventional wisdom. The present paper has proposed a theoretical justification for this phenomenon.

First I have noted that the basic economic assumptions of rivalry and excludability are not applicable to information shared over the Internet. Once created, information is intrinsically not a scarce good, and therefore there is no a priori reason to restrict access to it. On the contrary, freely distributing information is likely to profit its creator, since it helps to improve the quality of the information, and to enhance the creator's expertise and reputation. Moreover, open access obviously profits everybody else, and in particular those who otherwise would be too poor to pay for the information.

I then used the paradigm of self-organization through stigmergy to explain how open access development can be efficiently coordinated. Thanks to websites listing "work-in-progress", people willing to contribute to the collective development of an information product are efficiently directed to the tasks where their contribution is most likely to be fruitful. This obviates the need both for centralized planning and control, and for the "invisible hand" of the market matching supply to demand.

These innovations appear fundamental enough to revolutionize our socio-economic system (cf. Weber, 2004), offering high hopes for the future, e.g. in stimulating innovation, education, democratization, and economic development. While open-access distribution is not applicable to material resources, their cost as a fraction of the total economic cost of any good or service is becoming progressively smaller in a society that is ever more heavily dependent on information. Therefore, it could be theoretically envisaged that most economic value would eventually be produced under an open-access system. To make such a scenario less speculative, we will first need to investigate the complex issue of information production that requires considerable material investment, such as pharmaceutical research with its expensive equipment, where patents and other ways of "closing off" information are rife. The issue becomes less daunting, though, if we remember that this kind of research mostly builds on publicly funded (and thus normally open access) work.

To be able to fully compete with the established market-based system, moreover, the still very young open access movement will need to further learn from its experiences, addressing its remaining weaknesses and building further on its strengths. This will in particular require developing better standards and rules, and more powerful software solutions for harnessing stigmergy and allocating recognition and feedback—the main drivers behind the success of open access according to the present analysis.

For example, in the Wikipedia system—which otherwise keeps a very detailed track of all changes made to all documents by all users—it is impossible at present to get an overview of how much a particular user has contributed to the system. Given Wikipedia's versioning system, it should be possible to measure how much of the text entered by a given user survives in the present state of the encyclopedia. This would provide a useful measure of both the quantity and the quality of that author's contributions, thus establishing a benchmark by which to measure expertise and activity level. Similarly, more advanced algorithms (e.g. inspired by Google's PageRank or Hebbian learning, Heylighen, 1999) could be implemented to organize and prioritize tasks. Such intelligent methods for coordinating distributed information production could turn the World-Wide Web from merely a collective memory or shared workspace into a true "Global Brain" for humanity, that would be able to efficiently solve any problem, however complex (Heylighen, 1999, 2004).